\documentclass[5p,twocolumn,times,number]{elsarticle}

\usepackage{latexsym,amsmath,amssymb,theorem,epsfig}
\usepackage{graphicx}
\usepackage{bm}

\def\be{\begin{equation}}
\def\ee{\end{equation}}
\def\ba{\begin{eqnarray}}
\def\ea{\end{eqnarray}}
\def\({\left(}
\def\){\right)}

\def\nn{\nonumber}

\def\d{\mathrm{d}}
\def\mn{_{\mu \nu}}
\def\ab{_{\alpha \beta}}
\def\mupn{^\mu_{\, \nu}}
\def\mf{M_4^2}

\def\o{\omega}
\def\ph{\varphi}
\def\stu{St\"uckelberg }
\def\dmn{\partial_\mu\partial_\nu}

\def\p{\partial}

\begin{document}

\title{Massive gravity from Dirichlet boundary conditions}

\author[Mc,PI]{Claudia de Rham}
\ead{crham@perimeterinstitute.ca}
\address[Mc]{Department of Physics \& Astronomy, McMaster University, \\ Hamilton ON, L8S 4M1, Canada}
\address[PI]{Perimeter Institute for Theoretical Physics,
31 Caroline St. N., \\ Waterloo, ON, N2L 2Y5, Canada}

%\affiliation{
%$^{1}$Department of Physics \& Astronomy, McMaster University, Hamilton ON, L8S 4M1, Canada\\
%$^2$Perimeter Institute for Theoretical Physics,
%31 Caroline St. N., Waterloo, ON, N2L 2Y5, Canada
%}

\begin{abstract}
We propose an explicit non-linear realization of massive gravity, which
relies on the introduction of a spurious compact
extra dimension, on which we impose half-Newmann and half-Dirichlet boundary conditions.
At the linearized level, we recover the expected gravitational exchange
amplitude between two sources mediated by a massive Fierz-Pauli spin-2 field, while
cubic interactions in the additional helicity-0 mode give rise to
the expected Vainsthein mechanism. We also show that this framework can accommodate for a flat four-dimensional geometry in the presence of a cosmological constant, putting this framework on a good footing for the study of degravitation.
\end{abstract}


\maketitle

%%%%%%%%%%%%%%%%%%%%%%%%%%%%%%%%%%%%%%%%%%%%%%%%%%%%%%%%%%%%%%%%%%%%%%%%%%%%%%%%%%%%%%%%%%%%%%%%%%%%

\section{Introduction}

While laboratory experiments, solar systems tests and cosmological
observations have all been in complete agreement with General Relativity
for now almost a century, these bounds do not eliminate the
possibility for the graviton to bear a small hard mass $m\lesssim
6.10^{-32}$eV, \cite{Goldhaber:2008xy}. Conversely,
the main obstacle in giving the graviton a mass lies in the
theoretical constraints rather than the observational ones, as
explicit non-linear realizations of massive gravity are
hard to construct.
The Dvali-Gabadadze-Porrati (DGP) model is the first realization of soft massive gravity, where
the graviton can be thought of as a resonance, or a superposition of massive
modes \cite{DGP}. This model was then extended to higher
dimensions,  \cite{Gabadadze:2003ck,cascade}, where gravity becomes
even weaker at large distances, and could exhibit a ``degravitation"
mechanism, by which the cosmological constant could be large but
gravitate weakly on the geometry \cite{degravitation}.
Such a degravitation mechanism is also ``expected" to be present if
the graviton bears a hard mass. An explicit
realization of a theory of a hard mass gravity was proposed in \cite{Gabadadze:2009ja}, which appeared
while this work was in progress, and relies on the same mechanism.

This framework is based on the presence of a ``spurious" compact extra dimension on which
we impose Dirichlet boundary condition on one end and Neumann (Isra\"el) on the
other, where our 4d world stands. 
The techniques used throughout this study, in particular the introduction of a \stu field to restore 4d gauge invariance, are in no way original to this work, however the introduction on the spurious extra dimension provides a geometrical interpretation of massive gravity, for which non-linearities can be tracked down explicitly. Furthermore, this model is of high interest for the study of degravitation, providing a framework where explicit solutions with a cosmological constant can be understood and more general cosmological solution can be studied numerically.

We also show that when diffeomorphism is broken along the extra
dimension, one recovers an effective 4d theory of
gravity where the graviton has a constant mass.
Moreover, this class of model can
also accommodate a fully 5d diffeomorphism invariant theory for
which the 4d effective graviton has a soft mass and is free of any
ghost-like instability at the non-linear level. 

We proceed as follows: We first show in section \ref{sec:Scalar} how our mechanism works for a
scalar field toy-model before presenting the full spin-2 analogue in section \ref{sec:Gravity}.
We then recover the expected gravitational exchange amplitude
between two conserved sources for a theory of massive gravity in section \ref{sec:EffAction} and
derive the decoupling limit for a specific class of models where higher extrinsic curvature terms are
present in section \ref{sec:Decoupling}, while the decoupling limit in the more general case is
deferred for later studies. We also discuss on the number of physical degrees of freedom and comment on the stability (presence of ghosts) in this class of model. 
We then present in section \ref{sec:FlatSol} solutions capable of ``hiding" a 4d
cosmological constant by curving the extra dimension and keeping the
3-brane flat, which is of high importance for the degravitation mechanism. 
Finally, we discuss soft massive gravity in the appendix \ref{sec:Appendix}, which is obtained when restoring 5d gauge invariance along the extra dimension.  

%%%%%%%%%%%%%%%%%%%%%%%%%%%%%%%%%%%%%%%%%%%%%%%%%%%%%%%%%%%%%%%%%%%%%%%%%%%%%%%%%%%%%%%%%%%%%%%%%%%%
\section{Scalar Field Toy model}
\label{sec:Scalar}
Before diving into the technical subtleties of the
gravitational case, we focus to start with on the core of the idea using
a scalar field toy-model.
Let $\ph(x^\mu,\o)$ be a massless scalar field living in a 5d space-time $(x^\mu, \o)$ where
the coordinates $x^\mu$, $\mu=0,\cdots,3$ describe our four transverse dimensions, while the fifth
coordinate $\o$ is compact, $0\le\o\le \bar \o$, and we choose the $\o$ coordinate to be dimensionless.
We explicitly break the 5d Lorentz invariance by omitting the kinetic term along the transverse direction in the bulk
\ba
S= \int_0^{\bar \o}\d \o \, \d^4x \(\frac{M_5^4}{2} (\partial_\o\ph)^2 +\delta(\bar \o-\o)\mathcal{L}_4 \) \,,
\ea
while these kinetic terms are present on the brane:
\ba
\mathcal{L}_4=-\frac{M_4^2}{2} \ph \Box \ph+\ph J(x)\,,
\ea
where $\Box$ is the 4d d'Alembertian and
$J$ the source localized on the 3-brane.
A shift in the brane position $\bar \o$ is equivalent to rescaling the 5d scale $M_5$ and without loss of generality,
we set $\bar \o\equiv1$ and $M_5^4=M_4^2 m^2$, where $M_4$ is the 4d Planck scale and $m$ is a mass parameter.
The boundary condition on the brane at $\o=1$ is set using the standard Neumann or Isra\"el Matching
Conditions, while at $\o=0$, we choose to impose the Dirichlet boundary condition:
\ba
\ph(x,\o)\big|_{\,\o=0}&=&0\\
-\mf m^2 \partial_\o \ph \big|_{\, \o=1}&=&-\mf \Box \ph+J\,.
\ea
Solving the bulk equation of motion with the previous boundary condition, the field profile is therefore
$ \ph(x,\o)=\bar \ph(x) \o $, where $\bar \ph$ is the induced value of the field on the brane, satisfying the 4d effective
equation of motion on the brane,
\ba
\mf(\Box-m^2)\bar \ph=J(x)
\ea
and hence behaving as a massive scalar field from a 4d
point of view. Needless to say that this is a very convoluted way to
obtain a massive scalar field theory, but for gravity, it would be
extremely difficult to do so otherwise.

%%%%%%%%%%%%%%%%%%%%%%%%%%%%%%%%%%%%%%%%%%%%%%%%%%%%%%%%%%%%%%%%%%%%%%%%%%%%%%%%%%%%%%%%%%%%%%%%%%%%

\section{Massive Gravity}
\label{sec:Gravity}
The extension of this model to a spin-2
field is straight-forward. We consider a 4d metric $q\mn(x^\mu,\o)$ living
in the previous 5d space-time. 5d diffeomorphism is here again explicitly broken,
but 4d gauge invariance is preserved using the standard trick of introducing a St\"uckelberg
field $M^\mu$ with $N^\mu(x,\o)=\p_\o M^\mu(x,\o)$,
which shifts under a 4d gauge transformation $x^\mu  \to  \tilde x^\mu(x,\o)$ as
\ba
q\mn &\to& \tilde q\mn=q\ab \frac{\partial  x^\alpha}{\partial \tilde x^\mu} \frac{\partial  x^\beta}{\partial \tilde x^\nu}\,,\\
N^\mu &\to& \tilde N^\mu=N^\alpha\frac{\partial \tilde x^\mu}{\partial x^\alpha}+\partial_\o \tilde x^\mu
\ea
so that the ``extrinsic curvature"
\ba
K\mn=\frac 12 \mathcal{L}_n q\mn=\frac 12 \(\partial_\o q\mn-D_{(\mu}N_{\nu)}\)
\ea
transform as 4d tensor. Hereafter, the 4d metric $q\mn$ is used to express the covariant derivative $D_\mu$ as well as
to raise and lower the indices.

Similarly as for the scalar field toy-model, we then construct the 5d bulk action by
considering the equivalent of the
``5d curvature" $R_{5}[q,M]=R_4[q]+K^2-K\mupn K^\nu_{\, \mu}$
but omitting the contribution from the 4d kinetic term $R_4$
\ba
\label{toymodel1}
S_K= \frac{\mf m^2}{2}\int_0^1 \d \o \, \d^4x\sqrt{-q}  \(K^2-K\mupn K^\nu_{\, \mu}\)\,.
\ea
Notice that the specific combination $K^2-K\mupn K^\nu_{\, \mu}$ that appears when expressing the 5d curvature in terms of the 4d one, is precisely what will give to the specific Fierz-Pauli combination, which is the only ghost-free linear realization of massive gravity that respects 4d diffeomorphism invariance.
The 4d curvature is yet present on the brane at $\o=1$ which holds the action
\ba
S_4= \int \d^4x\sqrt{-q}  \(\frac{\mf}{2}R_4-\mathcal{L}_4\)\,,
\ea
where $\mathcal{L}_4$ is the Lagrangian for matter fields confined to the 3-brane.
Working in terms of the two dynamical variables $q\mn$ and $M^\mu$,
the Isra\"el matching conditions
are used to determine the boundary condition on the brane at $\o=1$, while we impose Dirichlet boundary condition at $\o=0$:
\ba
q\mn(x^\alpha,\o)\big|_{\, \o=0}\equiv \eta\mn\ \ {\rm and}\ \ M^\mu\big|_{\, \o=0}\equiv 0\,.
\ea
Notice that if we had restricted ourselves to theories that only
have the restricted gauge symmetry $x^\mu  \to  \tilde x^\mu(x)$,
the action \eqref{toymodel1} would be
gauge invariant without the need of the \stu field, but the
Dirichlet boundary condition would break 4d gauge invariance.
The extended symmetry $x^\mu  \to  \tilde x^\mu(x,\o)$ and the \stu
field therefore  play a crucial role.

Differentiating the bulk action with respect to the \stu field yields the ``Codacci" equation
\ba
D_\mu K\mupn-\p_\nu K=0\,,
\ea
while differentiating the action with respect to the metric leads to the modified ``Gauss" equation:
\ba
&&\hspace{-15pt}\mf
m^2\left\{\hspace{-2pt}\mathcal{L}_n\hspace{-2pt}\(K\mupn-K\delta\mupn\)+K K\mupn
-\frac 12 \(K^2+K^\alpha_{\, \beta}K^\beta_{\, \alpha}\)\hspace{-2pt}\delta\mupn\right\}\nn\\
&&\hspace{10pt}=\delta(\o-1)\(T\mupn-\mf G^{(4)}{}\mupn\)
\ea
where the Lie derivative of a $(1,1)$-tensor is
\ba
\mathcal{L}_nF\mupn=\(\p_\o-N^\alpha\p_\alpha\)F\mupn+F^\alpha_{\, \nu} \p_\alpha N^\mu-F^\mu_{\, \alpha}\p_\nu N^\alpha\,.\
\ea
In the absence of any gravitational source $\mathcal{L}_4$, the field $M^\mu$ vanishes
and the 4d metric is flat $q\mn=\eta\mn$ as in standard general relativity.
In what follows, we show that, this theory behaves
as a theory of massive gravity at the linear level.

\section{Effective Boundary Action}
\label{sec:EffAction}
We derive in this section the effective action on the 3-brane for small perturbations around the vacuum solution,
$q\mn=\eta\mn+h\mn(x,\o)$, sourced by a 4d stress-energy tensor $T\mn$ localized on the brane at $\o=1$. We follow the same approach as that used in \cite{Luty:2003vm}.
In terms of the variable $H\mn$,
\ba
H\mn=  h\mn-\p_{(\mu}M_{\nu)}=h\mn-(\p_\mu M_\nu+\p_\nu M_\mu)\,,
\ea
the bulk action is then of the form
\ba
\mathcal{L}_K=-\frac{\mf m^2}{8} \ \p_\o H^{\mu \nu}\, \p_\o
(H\mn-H_4\eta\mn)\,,
\ea
where $H_4=H^\alpha_\alpha$.
The field $H\mn$ is  hence linear in the fifth variable $\o$, and the Dirichlet boundary condition at $\o=0$ sets
\ba
H\mn (x^\mu,\o)= \bar H\mn(x^\mu)\, \o\,,
\ea
where hereafter bar quantities represent the induced value of the fields on the brane. Using this expression in $\mathcal L_K$, leads after integration by part to the 4d boundary term at $\o=1$:
\ba
\mathcal{L}^{\rm bdy}_{K}=-\frac{\mf m^2}{8} \ \bar H^{\mu \nu}\,  (\bar H\mn-\bar H_4 \eta\mn)\,,
\ea
which is precisely the mass term of a standard Fierz-Pauli massive theory of gravity at the linearized level. 
To this induced boundary action, we add the brane Einstein-Hilbert term
\ba
\mathcal{L}^{\rm bdy}_{R_4}\hspace{-3pt}=\hspace{-3pt}\frac{\mf}{8} \Big[\bar h^{\mu \nu}\Box (\bar h\mn-\bar h_4\eta\mn)
+2(\partial_\mu \bar h\mupn)^2+\bar h_4 \partial_\mu\partial_\nu \bar h^{\mu\nu} \Big]\nn \,,
\ea
which provides the kinetic for the massive Fierz-Pauli graviton. 
Since both boundary actions are invariant under the gauge transformation $x^\mu\to x^\mu+\xi^\mu(x) \o$, one can fix this gauge freedom by adding a gauge fixing term similarly as in \cite{Luty:2003vm},
\ba
\label{gf}
\mathcal{L}^{\rm bdy}_{\rm gf}=-\frac{\mf}{4} \,
\big(\p_\alpha\bar h^\alpha_{\, \mu}-\frac 12 \p_\mu \bar h_4- m^2 \bar M_\mu\big)^2 \,.
\ea
The resulting boundary action is then
\ba
&&\hspace{-18pt}\mathcal{L}^{\rm bdy}_{\rm eff}=\frac{\mf}{8}\Big[
\bar h^{\mu\nu}(\Box-m^2)\bar h\mn-\frac 12 \bar h_4 (\Box-2m^2)\bar h_4\ \ \\
&&\hspace{-5pt}+m^2\(F\mn^2+ \bar h_4 \p_\mu \bar M^{\mu}
+2m^2 \bar M_\mu^2\)\!
\Big]+\frac 12 \bar h\mn T^{\mu \nu}\nn\,,
\ea
with $F\mn=\p_{(\mu} M_{\nu)}$, and the second line corresponds to
the action of a Proca field coupled to $\bar h\mn$. Notice that in the absence of this coupling, the Proca or \stu field would be irrelevant.
When coupling these fields to conserved matter, only the scalar mode in the \stu field is excited, and the
resulting gravitational exchange amplitude between two sources is then
\ba
\mathcal A\sim-\frac 2 \mf\int \d ^4 x \, T'^{\mu\nu}\frac{1}{\Box-m^2}\(T\mn-\frac 13 T\eta\mn\)\,,
\ea
corresponding to the expected gravitational exchange amplitude due to a massive graviton.
In particular, we notice the standard factor $1/3\, T$ instead of $1/2\, T$ which
appears in massive gravity and signals the presence of an extra
helicity-0 mode hidden in the \stu field.
As observed by van Dam-Veltman and Zakharov (vDVZ), this factor remains $1/3$
even in the massless limit and is at the origin of the well-know
vDVZ discontinuity, \cite{vanDam:1970vg}. The
resolution to this puzzle lies in the observation that close enough to any source,
the extra scalar mode is strongly
coupled, \cite{Vainshtein:1972sx}. Non-linearities dominate over the linear term and
effectively freeze the field. This is most easily understood by
studying the decoupling limit.

%%%%%%%%%%%%%%%%%%%%%%%%%%%%%%%%%%%%%%%%%%%%%%%%%%%%%%%%%%%%%%%%%%%%%%%%%%%%%%%%%%%%%%%%%%%%%%%%%%%%
\section{Decoupling limit}
\label{sec:Decoupling}
Following the same prescription as in \cite{Luty:2003vm,ArkaniHamed:2002sp}, we
work from now on in the high energy limit $\Box\gg m^2$, and focus on the scalar
mode, $\bar M_\mu = -\partial_\mu \pi$. The helicity-0 mode then decouples when changing variable to $h'\mn=\bar h\mn+m^2 \pi \eta\mn$, and the effective boundary action simplifies to
\ba
\label{effactiondec}
\mathcal{L}^{\rm bdy}\simeq \frac{\mf}{4}\Big[\frac 12 h'^{\mu\nu} \Box ( h'\mn-\frac 12   h'_4 \eta\mn)
+3 m^4 \pi \Box \pi\Big]\,.
% \\ +\frac 12 (\bar h'\mn+m^2 \pi \eta\mn)T^{\mu \nu}\nn\,.
\ea
The small kinetic term of $\pi$ is precisely what resolves the vDVZ
discontinuity, similarly as in DGP, \cite{Luty:2003vm}.
In the small mass limit, higher order interactions in $\pi$ dominate over the quadratic term and effectively freeze the extra excitations out.
To see the strong coupling at work, let us
find out the most important interaction present beyond this quadratic action. We work for that in terms of the canonically normalized variables $\hat h\mn=M_4 h'\mn$ and $\hat \pi=m^2 M_4\pi$. A general bulk interaction between $\pi$ and $h'\mn$ will give rise to a boundary term of the form
\ba
\label{interactions}
\mathcal{L}_{{\rm bdy}}^{(p,q)}\sim\mf m^2 \(\frac{\partial^2 \hat \pi}{m^2 M_4}\)^q\(\frac{\hat h \mn}{M_4}\)^p\,.
\ea
We immediately see that interactions with the helicity-2 mode $h'\mn$ bear an important coupling scale and will hence be suppressed. Setting $p=0$, the strong coupling scale for this kind of interaction is
\ba
\Lambda_{q}\sim M_4 \(\frac{m}{M_4}\)^{\frac{2-2q}{4-3q}}\,.
\ea
The lowest interaction scale therefore occurs for cubic interactions $q=3$, as expected from \cite{ArkaniHamed:2002sp},
giving rise to the strong coupling scale
\ba
\Lambda_5=\(m^4 M_4\)^{1/5}\,.
\ea
We can quickly convince ourselves that such cubic
interactions generically exist in a theory of massive gravity, although they are absent in the specific theory at
hand as the \stu field $M^\mu=-\p_\mu \hat \pi /m^2 M_4$ only comes
to quadratic order in the action. For the cubic interactions with
scale $\Lambda_5$ to be present, the action should include cubic
terms in the \stu field such as $(\p_\mu M_\nu)^3$ not present in
the model considered thus far. However such terms will typically be
present if higher order terms in the extrinsic curvature are
present.

%%%%%%%%%%%%%%%%%%%%%%%%%%%%%%%%%%%%%%%%%%%%%%%%%%%%%%%%%%%%%%%%%%%%%%%%%%%%%%%%%%%%%%%%%%%%%%%%%%%%

\subsection{In the presence of $K^3$ terms}
Generically we expect to generate higher order in the extrinsic curvature by quantum
interactions, without modifying the linearized arguments provided so far.
 Such interactions are typically be of the form
\ba
\label{toymodel2}
\tilde{\mathcal{L}}_{K^3}= \frac{\mf m^2}{2}
\(\alpha K^3-(\alpha+\beta)K K\mn^2+\beta K\mn^3\)\,,
\ea
with $\alpha$ and $\beta$ arbitrary dimensionless
parameters.
We focus on the scalar mode $h\mn=m^2 \Pi
\eta\mn$ and $M_\mu=-\partial_\mu \Pi$, which extends in the bulk as $\Pi=\pi(x)\,
\o$.
At high energy, these terms contribute to the boundary action with
the following cubic interactions
\ba
\label{L3}
\hspace{-15pt}\tilde{\mathcal{L}}_{\rm K^3}^{(3)}=\frac{1}{2\Lambda_5^5} \Big(
\alpha (\Box \hat \pi)^3-(\alpha+\beta) (\Box \hat \pi)(\dmn \hat \pi)^2+\beta(\dmn \hat \pi)^3\Big)
\ea
which dominate over the quadratic term at the energy scale $\Lambda_5$.
% Even though quadratic terms in the extrinsic curvature in \eqref{toymodel1} can in principle cancel these interactions, this would only happen for a specific value of $\alpha$ and $\beta$, and such terms are therefore quite generic.
These cubic interactions are precisely the ones expected for a
typical theory of massive gravity in \cite{ArkaniHamed:2002sp,degravitation}, and are the ones responsible for
the Vainsthein mechanism, \cite{Vainshtein:1972sx}.
At least in the decoupling limit, this theory exhibits the
degravitation behavior \cite{degravitation} and therefore represents
a powerful tool to study this mechanism further in a fully non-linear
scenario.

%%%%%%%%%%%%%%%%%%%%%%%%%%%%%%%%%%%%%%%%%%%%%%%%%%%%%%%%%%%%%%%%%%%%%%%%%%%%%%%%%%%%%%%%%%%%%%%%%%%%

\subsection{General $K^n$ terms}

In more generality, one may expect the extrinsic curvature interactions to come in at the order $n\ge2$. They will then generate interactions of the form
\ba
\label{toymodeln}
\tilde{\mathcal{L}}_{K^n}\sim \mf m^2
\(\frac{\Box \hat{\pi}}{M_4 m^2}\)^n\sim \frac{1}{\Lambda_{\star}^{3n-4}}(\Box \hat{\pi})^n\,,
\ea
with the strong coupling scale
\ba
\Lambda_\star=\(m^{\frac{2(n-1)}{n-2}}M_4\)^{\frac{n-2}{3n-4}}\,,
\ea
in particular we recover the strong coupling scale $\Lambda_\star=\Lambda_5=\(m^4M_4\)^{1/5 }$, when extrinsic curvature interactions are included already at cubic order ($n=3$), whereas if the theory is free of such interactions or $n\to \infty$, the strong coupling scale is $\Lambda_\star=\Lambda_3=\(m^2M_4\)^{1/3 }$.

%%%%%%%%%%%%%%%%%%%%%%%%%%%%%%%%%%%%%%%%%%%%%%%%%%%%%%%%%%%%%%%%%%%%%%%%%%%%%%%%%%%%%%%%%%%%%%%%%%%%

\subsection{Ghosts and physical degrees of freedom}

As soon as higher extrinsic curvature terms are present, they result in interactions that are relevant at the scale
$\Lambda_\star$. In that case, the theory inexorably manifests a ghost at the non-linear
order. This can be seen by the presence of the higher order
derivative operators of the form ``$\Box^2 \pi$" that appears in
the equation of motion of \eqref{L3} or \eqref{toymodeln}.
Such a ghost is expected in a theory
of hard mass gravity, and is usually refereed to as the Boulware-Deser ghost, \cite{Boulware:1973my, Creminelli:2005qk}.
This theory has $10$ degrees in the metric and $4$ in the \stu
field, but the gauge invariance makes only 6 of them physical, like in a usual theory of massive gravity around a general background. The \stu field contributes with 4 additional degrees of freedom, compared to the only 2 present in a theory of massless gravity.

However, when perturbing to first order around flat space-time, 
only 5 degrees of freedom are excited, ($M_\mu$ plays the role of a
Proca field, with only 3 degrees of freedom, one of them being the helicity-0 mode $\pi$, while the two helicity-1 modes decouple when considering conserved sources), as expected from a usual Fierz-Pauli massive theory of gravity.
At the non-linear level, the 6th mode is typically excited and propagates a ghost, at least when higher extrinsic curvature terms are present, or in other words when the strong coupling scale is below $\Lambda_3$.

%%%%%%%%%%%%%%%%%%%%%%%%%%%%%%%%%%%%%%%%%%%%%%%%%%%%%%%%%%%%%%%%%%%%%%%%%%%%%%%%%%%%%%%%%%%%%%%%%%%%

\subsection{In the absence of higher extrinsic curvature terms}
We emphasize however that when the theory is exempted of any higher extrinsic curvature term $K^n$ (with $n>2$),
all interactions with coupling scale $\Lambda_q$ with $ 1/5 \le q<1/3$ disappear.
Indeed, in that case interactions of the form \eqref{interactions} are only possible with $0\le q\le2$, since the \stu field only comes in at quadratic order in the action.  In that case the associated strong coupling scale is then $\Lambda_3=(m^2 M_4)^{1/3}$, and interactions becoming important at that scale can be of any order.
The situation is then far more subtle. In particular, it has been shown in
\cite{Gabadadze:2009ja} that the Hamiltonian density remains
positive definite for appropriate choice of boundary conditions when
these $K^n$ terms are absent. 
Furthermore, the strong coupling scale in this case is the same as in the DGP model \cite{DGP} (or its extension in the appendix), for which no ghost-like instability is manifest non-linearly.
Understanding whether the theory \eqref{toymodel1} has an underlying symmetry that
keeps only 5 physical degrees of freedom non-linearly, or in other words whether
or not the Boulware-Deser ghost manifests itself in that case
 and if so at which scale therefore deserves more attention and will
 presented in some later work, \cite{next}.
Before concluding we show that this model
of massive gravity can be of great interest for cosmology as it can
accommodate for flat solutions in the presence of a cosmological constant on the brane.

%%%%%%%%%%%%%%%%%%%%%%%%%%%%%%%%%%%%%%%%%%%%%%%%%%%%%%%%%%%%%%%%%%%%%%%%%%%%%%%%%%%%%%%%%%%%%%%%%%%%

\section{Flat Solutions with Tension}
\label{sec:FlatSol}
We show here that such models present solutions which are very
similar to the codimension-2 ``deficit-angle" configuration, that carry a
tension but keep the 4d geometry flat. Indeed, including a
cosmological constant $\lambda_4$ on the brane, gives rise
to the following metric profile: $q\mn=a^2(\o)\eta\mn$ with
\ba
a^2(\o)=\frac{\o+\o_0}{\o_0}\,,
\ea
where $\o_0$ is a positive constant, related to $\lambda_4$ via the
Isra\"el Matching condition:
\ba
\lambda_4=\frac32 \frac{m^2 M_4^2}{1+\o_0}\,.
\ea
The 4d geometry on the brane is flat, and the cosmological constant
on the brane is carried by the bulk profile of the metric. Notice
however, that similarly to the deficit-angle case for codimension-2
branes, such solutions only exist when the tension is smaller than a maximal value $\lambda_{max}=\frac 32 m^2
\mf$. When higher order terms in the extrinsic curvature are
included, this bound can be increased slightly but not by a significant
order of magnitude.

%%%%%%%%%%%%%%%%%%%%%%%%%%%%%%%%%%%%%%%%%%%%%%%%%%%%%%%%%%%%%%%%%%%%%%%%%%%%%%%%%%%%%%%%%%%%%%%%%%%%
\section{Discussion}
\label{sec:Discussion}
In this paper, we constructed a class of models, first presented in \cite{Gabadadze:2009ja},
giving rise to theories of massive gravity, with hard or soft
masses depending on the details of the setup. This model relies on
the presence of a spurious compactified extra dimension on which we
impose half-Neumann, half-Dirichlet boundary conditions. For
definiteness, we have focused most of this paper on a theory of
massive gravity with a constant mass, for which 5d diffeomorphism is
broken, and refer to the appendix for other kinds of solutions. In
the case of a hard mass, we recover the usual decoupling limit with strong coupling scale $(m^4 M_4)^{1/5}\le \Lambda_\star<(m^2 M_4)^{1/3}$
and show the presence of the Boulware-Deser ghost {\it when higher order terms in the extrinsic curvature are
considered}. When such terms are absent, all
the interactions with coupling scale $\Lambda_\star<(m^2 M_4)^{1/3}$ disappear and the
decoupling limit is more subtle.
In particular, using a Hamiltonian approach, it has been shown in
\cite{Gabadadze:2009ja}, that the energy is positive for appropriate
choices of boundary conditions when these $K^n$ terms are absent. It
will therefore be interesting to understand this result in the decoupling limit, \cite{next}.
In parallel, this model allows us to understand in more depth several
aspects of massive gravity and degravitation.
In particular, this model would allow us to understand how strong coupling explicitly
works in the case of a spherically symmetric source both when the
higher order interaction terms are present and the expected decoupling
limit is recovered, similarly as in \cite{Babichev:2009jt},
as well as in the more interesting case where these higher order extrinsic curvature terms are
absent.
Furthermore, independently of the presence or not of ghosts, this framework will allow us to
understand whether a theory of massive gravity
continues to exhibit the degravitation behavior at the fully non-linear
level and whether it can represent a successful tool to tackle the
cosmological constant problem.
It particular, we have shown the presence of static solutions in the
presence of a cosmological constant, and one should understand
whether or not such solutions are late-time attractors. Finally,
such solutions can only carry a maximal tension, and one
should understand whether this framework can be extended to
accommodate larger tensions, similarly as in \cite{deRham:2009wb}.

%%%%%%%%%%%%%%%%%%%%%%%%%%%%%%%%%%%%%%%%%%%%%%%%%%%%%%%%%%%%%%%%%%%%%%%%%%%%%%%%%%%%%%%%%%%%%%%%%%%%
\section*{Acknowledgments} I am extremely grateful to G.~Gabadadze,  J.~Khoury and A.~J.~Tolley for
very fruitful discussions. This work was supported in part at the Perimeter
Institute by NSERC and Ontario's MRI.

%%%%%%%%%%%%%%%%%%%%%%%%%%%%%%%%%%%%%%%%%%%%%%%%%%%%%%%%%%%%%%%%%%%%%%%%%%%%%%%%%%%%%%%%%%%%%%%%%%%%
\section*{Appendix: Soft Massive gravity}
\label{sec:Appendix}
To finish, we show in this appendix that when 5d Lorentz invariance is restored, the graviton acquires a soft mass, similarly as in DGP or Cascading Gravity, and is free of Boulware-Deser ghost instabilities.
Indeed, when considering the 5d diffeomorphism invariant action
\ba
S_5=\frac{M_5^2}{2}\int_0^{\bar y} \d y\d^4x\sqrt{-g_5} R_5\,,
\ea
working now in terms of the dimensionful direction $y$ which remains compactified. Imposing the Dirichlet
boundary condition at $y=0$ and the Neumann one at $y=\bar y$,  the metric perturbations satisfy the following bulk profile in 5d de Donder gauge,
\ba
h_{AB}(x,y)=\frac{\sinh (y \nabla)}{\sinh (\bar y \nabla)} \bar h_{AB}(x)
\ea
where $\nabla=\sqrt{-\Box}$. In terms of the graviton mass $m(\Box)$,
the gravitational exchange amplitude between two conserved sources at $\bar y$ is
\ba
\mathcal A\sim-\frac 2 \mf\int \d ^4 x \, T'^{\mu\nu}\frac{1}{\Box-m^2(\Box)}\(T\mn-\frac 13 T\eta\mn\)\,,
\ea
where the graviton mass is
\ba
m^2(\Box)=m_5\nabla \coth  (\bar y \nabla)\,,
\ea
with $m_5=M_5^3/\mf$. In particular, we recover the standard DGP behavior for large $\bar
y$,  while the opposite limit gives rise to a constant mass, similar to Cascading Gravity, \cite{cascade}
\ba
\bar y \nabla \gg 1,\  m^2 \to m_5\nabla \hspace{10pt}{\rm and}\hspace{10pt}
\bar y \nabla \ll 1,\  m^2 \to \frac{m_5}{\bar
y}\,.
\ea
Notice that the Dirichlet boundary condition at $y=0$ has projected
out the zero mode, and we do not recover 4d gravity in the infrared
limit, despite having a compactified extra dimension. Had we impose
the Neumann boundary conditions $\p_y h_{AB}|_0=0$ or periodic
boundary conditions,  $h_{AB}(0)=h_{AB}(\bar y)$, the zero mode
would survive and would be the dominant one in the infrared.

In this case, the decoupling limit arises precisely in the same way
as in DGP, \cite{Luty:2003vm}. The main difference with the model
presented in \eqref{toymodel1} is the presence of the lapse, which
plays a crucial role. The $\pi$ mode decouples at the strong
scale $\Lambda_3=(m_5^2 M_4)^{1/3}$ and its equation of motion
is then
\ba
3 \Box \hat \pi+\frac{1}{\Lambda_3^3}\((\Box \hat \pi)^2-(\p_\mu \p_\nu \hat
\pi)^2\)=-\frac{T}{M_4}\,.
\ea
As already hinted in this limit, where the 5d diffeomorphism is
restored, the theory is free of any ghost-like instability, when
working around the standard branch. However, similarly as the DGP model, this will not provide a satisfactory framework for degravitation, since it cannot accommodate for stable static solutions in the presence of a tension. 
We can check this statement explicitly, by deriving the effective Friedmann equation on the
brane. For that, we consider the bulk metric
\ba
\d s^2=\d y^2-\frac{1}{1+\kappa y}\, \d t^2+(1+\kappa y)\, \delta_{ij}\d x^i \d x^j\,,
\ea
where $\kappa$ is a free parameter, analogue to the spatial curvature
which can be scaled to 1. If the brane is located at $y=\bar y(t)$,
the induced extrinsic curvature on the brane is then
\ba
K_{ij}=\frac{\kappa}{2\sqrt{1-a^2(t)\, \bar y'{}^2}}\
\delta_{ij}\,,
\ea
where $a^2(t)=(1+\kappa\bar y(t))$, and the resulting Friedmann equation in the presence of a fluid
with energy density $\rho$ is
\ba
\mf\(3H^2+\frac{m_5}{2}\sqrt{4H^2+\frac{\kappa}{a^4}}\, \)=\rho\,.
\ea
When $\kappa/a^4\ll H^2$, we recover the intermediary regime analogue to
DGP, 
\ba
\mf(3H^2+m_5 H)=\rho\,,
\ea 
while in the opposite limit, the
corrections just play the role of a spatial curvature term. 

%%%%%%%%%%%%%%%%%%%%%%%%%%%%%%%%%%%%%%%%%%%%%%%%%%%%%%%%%%%%%%%%%%%%%%%%%%%%%%%%%%%%%%%%%%%%%%%%%%%%

\end{document}